\documentclass[aps,pra,showpacs,twocolumn,floatfix]{revtex4}

\usepackage{amsmath}
\usepackage{bm}
\usepackage{graphicx}

\begin{document}

\title{On the low-temperature diffusion of localized Frenkel excitons in
linear molecular aggregates}

\author{A.\ V.\ Malyshev}
\thanks{On leave from Ioffe Physiko-Technical Institute,
26 Politechnicheskaya str., 194021 Saint-Petersburg, Russia}
\affiliation{Departamento de F\'{\i}sica de Materiales, Universidad
Complutense, E-28040 Madrid, Spain}

\author{V.\ A.\ Malyshev}
\thanks{On leave from ``S.I. Vavilov State Optical Institute'',
Saint-Petersburg, Russia.}
\affiliation{Departamento de F\'{\i}sica de Materiales, Universidad
Complutense, E-28040 Madrid, Spain}

\author{F.\ Dom\'{\i}nguez-Adame}
\affiliation{Departamento de F\'{\i}sica de Materiales, Universidad
Complutense, E-28040 Madrid, Spain}

\date{\today}

\begin{abstract}

We study theoretically diffusion of one-dimensional Frenkel excitons in
J-aggregates at temperatures that are smaller or of the order of the J-band
width. We consider an aggregate as an open linear chain with uncorrelated
on-site (diagonal) disorder that localizes the exciton at chain segments of
size smaller than the full chain length. The exciton diffusion over the
localization segments is considered as incoherent hopping. The diffusion is
probed by the exciton fluorescence quenching which is due to the presence of
point traps in the aggregate. The rate equation for populations of the
localized exciton states is used to describe the exciton diffusion and
trapping. We show that there exist two regimes of the exciton diffusion at
low temperatures. The first, slower one, involves only the states of the
very tail of the density of states, while the second, much faster one, also
involves the higher states that are close to the bottom of the exciton band.
The activation energy for the first regime of diffusion is of the order of
one fifth of the J-band width, while for the second one it is of the order
of the full J-band width. We discuss also the experimental data on the fast
low-temperature exciton-exciton annihilation reported recently by I.\ G.\
Scheblykin {\em et al}, J. Phys. Chem. B {\bf 104}, 10949 (2000).

\end{abstract}


\maketitle

\section{Introduction}
\label{Intro}

Since the seminal works by Jelley~\cite{Jelley36} and
Scheibe~\cite{Scheibe36}, the concept of Frenkel
excitons~\cite{Frenkel31,Davydov71,Agranovich82} has been used for
explaining of the remarkable optical properties of molecular J-aggregates:
(i) the appearance of a narrow and intense line in the red-wing of the
absorption spectra (so called J-band), the full width of which is of the
order of several tens of wavenumbers at cryogenic temperatures and (ii) the
increase of the oscillator strength of the optical transition by almost two
orders of magnitude~\cite{deBoer90,Fidder90,Fidder91a,Fidder91b,Fidder93}.
During the nineties, a considerable progress in understanding of linear and
nonlinear optical dynamics of J-aggregates was made (for details see the
reviews~\onlinecite{Spano94,AdvMat95,Kobayashi96} and references therein).
In spite of the fact that monomers which form the aggregates have complex
chemical structure, both linear and nonlinear optical dynamics in
J-aggregates have been successfully described on the basis of the simplest
one-dimensional (1D) tight-binding model with diagonal and/or off-diagonal
disorder, both uncorrelated~\cite{Knapp84,Fidder91b,Knoester93} and
correlated~\cite{Knapp84,Durrant94,Malyshev99a,Malyshev99c}.

The eigenstates of a homogeneous (non-disordered) J-aggregate extend 
over the whole ($N$ monomers) aggregate. Disorder localizes
the lowest in energy exciton states at segments of about $N^*$ molecules;
$N^*$ depends on the disorder magnitude and is typically much smaller than 
the total number of molecules in the chain: $N^* \ll N$. One of the most
important consequences of this localization is the appearance of states
below the bottom of the bare exciton band. These states form the tail
of the density of states (DOS) and carry almost the whole oscillator
strength of the aggregate. For this reason the one-exciton absorption in
J-aggregates is spectrally located at the tail of the DOS (see, for
instance, Refs.~\onlinecite{Fidder91b,Fidder93}) and the width of the
absorption band is of the order of the width of the DOS tail.

The exciton diffusion in a disordered aggregate is essentially the
transition from one localized eigenstate to another.  The transition
probability depends particularly on the temperature, the energy spacing
between the involved states and the overlap of these states. The lower
states, being localized at {\it different} $N^*$-molecule segments of the
aggregate, overlap very weakly~\cite{Malyshev01b}. Contrary to that, the
higher exciton states, that are localized at segments larger than $N^*$
molecules, overlap strongly with several lower tail states. Although higher
states are thermally less favorable, the hops from the lower to higher
states can be faster than between the lower states because of the higher
overlap. In this paper, we show that this competition between the overlaps
and thermal favorability result in a complex scenario of the exciton
transport at low temperatures. At zero temperature an exciton resides in one
of the lower states at the tail of the DOS. As temperature rises, first, the
exciton starts to diffuse over the weakly overlapped states of the DOS tail.
The activation energy for this regime is of the order of $1/4$ of the DOS
tail width (that is of the order of the J-band width). The diffusion in this
regime is very slow. As the temperature increases further, the higher states
come into play. As these states overlap much better with the lower states
and each other and also are more extended, the diffusion rate increases by
several orders of magnitude. The activation energy for this faster regime of
the exciton diffusion is of the order of the DOS tail width or, in other
words, of the order the J-band width.

To the best of our knowledge, these aspects of the 1D diffusion problem have
not been discussed in the literature yet. The same tight-binding Hamiltonian
was used to describe transport properties of electrons in doped
semiconductors~\cite{Shklovskii84} as well as those of optical excitations
in activated glasses~\cite{Basiev87,Malyshev92}. It is to be stressed that
despite the seeming similarity of these problems the outlined scenario of
the low temperature 1D diffusion over the localization segments is more
complex than the diffusion over one-level point impurity centers in
semiconductors or glasses. The major complication comes from the fact that
the exciton can hop sideways to a different segment not only directly (a
"horizontal" hop) but also indirectly via higher states, so that "vertical"
hops up in energy become extremely important. In fact, it is the indirect
hops that provide the dominant contribution to the diffusion rate at
temperatures of the order of the J-band width.

We use the quenching of the exciton fluorescence by point traps to
probe the exciton diffusion. The temperature range lower or of the
order of magnitude of the J-band width is of our primary interest;
higher temperatures are beyond the scope of the present work.

The outline of the paper is as follows. In
Sec.~\ref{Model}, we present the microscopic model of exciton trapping.
Section~\ref{Qualit} is focused on the qualitative discussion of the
channels of the exciton diffusion over the localization segments. 
The results of our numerical simulations of the exciton
fluorescence quenching, obtained on the basis of the rate equation
approach, are the contents of Sec.~\ref{Numerics}. In
Sec.~\ref{Concl} we conclude the paper and discuss the results of
the recent experiments on the fast exciton-exciton annihilation in
the aggregates of the triethylthiacarbocyanine salt of
3,3'-bis(sulfopropyl)-5,5'-dichloro-9-ethylthiacarbocyanine
(THIATS)~\cite{Scheblykin00}.

\section{Microscopic model of the exciton fluorescence quenching}
\label{Model}

We model a J-aggregate by $N$ ($N \gg 1$) optically active
two-level molecules forming a regular in space 1D open chain. The
corresponding Frenkel exciton Hamiltonian reads~\cite{Davydov71}
(for the sake of simplicity only the nearest-neighbor interaction
is considered)
\begin{equation}
    H = \sum_{n=1}^{N}\> E_n |n\rangle \langle n|
    - J \sum_{n=1}^{N - 1}\> (|n + 1\rangle \langle n|
    + |n\rangle \langle n + 1| )\ .
    \label{H}
\end{equation}
Here $E_n$ is the excitation energy of the $n$-th molecule, $|n\rangle$ 
denotes the state vector of the $n$-th excited molecule.
The energies $E_n$ are assumed to be Gaussian uncorrelated
(for different sites) stochastic variables distributed around the
mean value $\omega_0$ (which is set to zero without loss of 
generality) with the standard deviation $\Delta$. The hopping
integral, $-J$, is considered to be non-random and negative ($J >
0$), which corresponds to the case of J-aggregates (see, e.g.,
Ref.~\cite{deBoer90}). In this case the states coupled to the
light are those close to the bottom of the exciton band. In what
follows, moderate disorder ($\Delta < J$) is considered. This
implies that the exciton eigenstates $\varphi_{\nu} \, (\nu = 1,
2, \ldots, N)$, found from the Schr\"oedinger equation
\begin{equation}
\sum_{m=1}^N H_{nm}\varphi_{\nu m} = \varepsilon_\nu \varphi_{\nu n} \ ,
\qquad
H_{nm}=\langle n|H|m \rangle\ ,
\label{k}
\end{equation}
are extended over relatively large segments of the chain. However,
the typical size of these localization segments, $N^*$, is small
compared to the full chain length $N$ (units of the lattice constant
are used throughout the paper).

Having been excited into an eigenstate $\nu$, an exciton cannot hop to
other eigenstates if coupling to vibrations is not taken into account. We
assume that this coupling is weak and do not consider polaron effects. This
limit is applicable to a number of J-aggregates as the Stokes shift of the
luminescence spectra with respect to the absorption spectra is usually
small~\cite{Fidder90,Fidder93}. The exciton-vibration interaction causes the
{\it incoherent} hopping of excitons from one eigenstate to another. We take
the hopping rate from the state $\nu$ to the state $\mu$ in the following
form (see, e.g., Ref.~\onlinecite{Leegwater97})
\begin{multline}
W_{\mu\nu} = W_0\ S(\left|\varepsilon_\nu -
\varepsilon_{\mu}\right|) \,\sum_{n=1}^N \varphi_{\nu n}^2
\varphi_{n\mu}^2 \, \\ \times
\begin{cases}
n(\varepsilon_\mu - \varepsilon_{\nu}), &\quad \varepsilon_\mu >
\varepsilon_{\nu}\\ 1+n(\varepsilon_{\nu} - \varepsilon_\mu),
&\quad \varepsilon_\mu < \varepsilon_{\nu}
\end{cases}\ .
\label{1Wkk'}
\end{multline}
Here, the constant $W_0$ characterizes the amplitude of hopping and
$n(\varepsilon) = [\exp(\varepsilon/T) - 1]^{-1}$ is the occupation number
of the vibration mode with the energy $\varepsilon$ (the Boltzmann constant
is set to unity). Due to the presence of the $n(\varepsilon)$ and
$1+n(\varepsilon)$ factors, the rate $W_{\mu\nu}$ meets the principle of
detailed balance: $W_{\mu\nu} = W_{\nu\mu}\exp[(\varepsilon_{\nu} -
\varepsilon_\mu)/T]$. Thus, in the absence of decay channels, the
eventual exciton distribution is the Boltzmann equilibrium distribution. The
sum over sites in (\ref{1Wkk'}) represents the overlap integral of exciton
probabilities for the states $\mu$ and $\nu$. The spectral factor
$S(|\varepsilon_\nu - \varepsilon_{\mu}|)$ depends on the details of the
exciton-phonon coupling as well as on the DOS of the medium
into which the aggregate is embedded. For example, within the Debye model
for the density of phonon states, this factor takes the form
$S(E_\nu - E_{\mu}) = (|E_\nu - E_{\mu}|/J)^3$~\cite{Bednarz02}. 
However, this model is applicable to glassy media (the media we assume as
the host) only in a narrow frequency interval of the order of several
wavenumbers (see, for instance, Ref.~\onlinecite{Dean72,Ovsyankin87}).
Therefore, as in Refs.~\cite{Shimizu01,Bednarz01}, we restrict ourselves to
a linear approximation to this factor, $S(E_\nu - E_{\mu}) = |E_\nu
- E_{\mu}|/J$. This accounts for reduction of the exciton-vibration
interaction in the long-wave acoustic
limit~\cite{Davydov71,Agranovich82}. Also, it eliminates the divergence of
$W_{\nu\mu}$ at small values of $|E_\nu - E_\mu|$.

The diffusion of Frenkel excitons can be probed by quenching of the exciton
fluorescence by traps. Consider an aggregate with point traps, namely
monomers at which an exciton decays non-radiatively and very fast compared
to the typical spontaneous emission rate of the aggregate. Then those
excitons that reach the traps decay non-radiatively and contribute to the
fluorescence quenching. If an exciton is created far from the trap it has to
diffuse to the trap to be quenched, the faster it diffuses the more
effective is the fluorescence quenching. Thus, the quenching rate depends on
the diffusion rate and can be used as a probe of the latter.

We define the quenching rate of the exciton state $\nu$ as:
\begin{equation}
\Gamma_\nu = \Gamma \sum_{i=1}^{N_q} |\varphi_{\nu i}|^2 \ ,
\label{Gamma}
\end{equation}
where $\Gamma$ is the amplitude of exciton quenching and the sum
runs over positions of the $N_q$ traps. Thus, we take the quenching rate
to be proportional to the probability to find the exciton at trap
sites.

We describe the process of the exciton trapping by means of the
rate equation:
\begin{equation}
{\dot P}_\nu = -(\gamma_\nu + \Gamma_\nu) P_\nu + \sum_{\mu = 1}^N
(W_{\nu\mu}\,P_{\mu} - W_{\mu\nu}\,P_\nu)\ , \label{P_nu}
\end{equation}
where $P_\nu$ is the population of the $\nu$th exciton eigenstate
and the dot denotes the time derivative, $\gamma_\nu =
\gamma\,f_\nu$ is the spontaneous emission rate of the $\nu$th
exciton state, while $\gamma$ is that of a monomer,
$f_\nu=(\sum_{n=1}^N \varphi_{\nu n})^2$ being the oscillator
strength of the state $\nu$.

The temperature dependence of the exciton quenching is calculated
as follows. We admit the definition of the exciton fluorescence
decay time as the integrated total population~\cite{Bednarz02}:
\begin{equation}
\tau = \int_0^\infty dt \, \sum_{\nu = 1}^N \langle P_\nu(t) \rangle \ ,
\label{tau}
\end{equation}
where angle brackets denote the average over disorder realizations
and traps positions. The decay time has to be calculated for aggregates
with traps (denoted as $\tau$) and without traps (denoted as
$\tau_0$). The quenching rate is then defined as
\begin{equation}
W_q = \frac{1}{\tau} - \left(\frac{1}{\tau}\right)_{N_q=0} \equiv
\frac{1}{\tau} -\frac{1}{\tau_0} \ .
\label{Wq}
\end{equation}
This quantity carries information about the diffusion rate and is
the object of our analysis.

The definition of the decay rate as the integrated total
population allows for considerable simplification of the
calculation procedure. Write the solution of Eq.~(\ref{P_nu})
in the formal matrix form
\begin{equation}
P_\nu(t) = \sum_{\mu=1}^N \left( e^{-{\hat R}t}
\right)_{\nu\mu}P_{\mu}(0) \ , \label{FormSol}
\end{equation}
where
\begin{equation}
R_ {\nu\mu} =
\left ( \gamma_\nu + \Gamma_\nu + \sum_{\mu=1}^N W_{\mu\nu}
\right ) \delta_{\mu\nu} - W_{\nu\mu} \ .
\label{R}
\end{equation}
After substitution of (\ref{FormSol}) into Eq.~(\ref{tau}) and
integration over time, $\tau$ can be expressed in terms of the
$\hat{R}$-matrix:
\begin{equation}
\tau =  \sum_{\nu,\mu=1}^N \left\langle {\hat R}^{-1}_{\nu\mu} \,
P_\mu(0) \right\rangle \ .
\label{1tau}
\end{equation}
Calculation of the quenching rate $W_q$ requires the calculation of the
inverse matrix ${\hat R}^{-1}$ for each realization of disorder
rather than the fluorescence kinetics. The inverse matrix is to be
found twice: for an aggregate with and without traps. Note that
the decay time $\tau_0$ also depends on temperature (see, for example,
Refs.~\cite{Bednarz02,Bednarz01,Spano90}).

\section{Qualitative arguments}
\label{Qualit}

At low temperatures excitons reside in the tail of the DOS, that is, below
the bottom of the bare exciton band, $E = -2J$. As we show below, higher
states that are close to the bottom of the bare band contribute to the
exciton diffusion as well. Therefore, these two parts of the exciton energy
spectrum are of primary importance for the low-temperature exciton
transport.

\subsection{Analyzing the low energy structure}
\label{Hidden}

\begin{figure}[ht!]
\includegraphics[width=\columnwidth,clip]{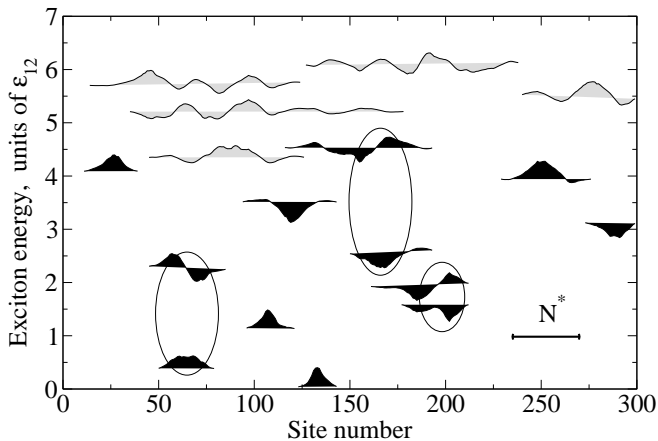}
\caption{The energy structure of the exciton states in the
vicinity of the bottom of the exciton band. The states are obtained by
diagonalization of the Hamiltonian~(\protect\ref{H}) for a linear chain of
300 molecules and the disorder magnitude $\Delta= 0.1J$. The baseline of
each state represents its energy in units of the spacing in the local energy
structure $\varepsilon_{12}$. The origin of the exciton energy is set to
the lowest energy for the realization. The wave functions are in arbitrary
units. It is clearly seen that some lower states can be grouped into local
manifolds (the manifolds are joined by ellipses). The states within each
manifold are localized at the same segment of typical length $N^*$
(this length is given by the bar in the lower right corner), they overlap
well with each other and overlap much weaker with the states localized at
other segments. The higher states (filled with gray color) are more extended
than the lower ones. Typically, they overlap well with several lower
states and with each other.}
\label{fig1}
\end{figure}

Here, we recall briefly the concept of the local (hidden) energy structure
of localized 1D excitons~\cite{Malyshev91,Malyshev95,Malyshev99a}, which was
proved to exist in the vicinity of the band
bottom~\cite{Malyshev01b,Malyshev01a}. According to this concept, the
low-energy one-exciton eigenfunctions obtained for a fixed realization of
disorder are localized at segments of typical size $N^*$ (localization
length). Some of these localized states (about 30\%) can be grouped into
local manifolds of two (or sometimes more) states that are localized at the
same $N^*$-molecule segment (see the states filled with black color and
joined by ellipses in Fig.~\ref{fig1}).
\begin{figure}[ht!]
\includegraphics[width=\columnwidth,clip]{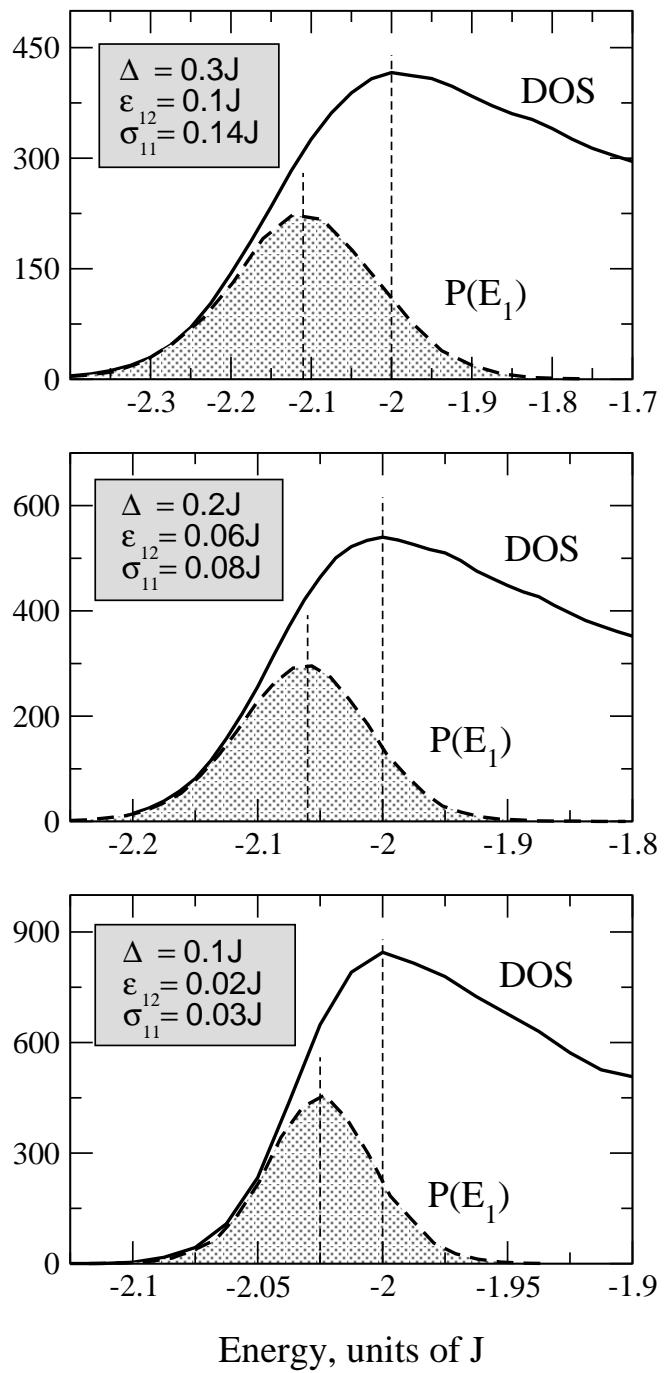}
\caption{The total DOS (solid line) and the DOS of the {\it local} 
ground states $P(E_1)$ (dashed line) for different magnitudes of disorder
$\Delta$. The DOS is normalized to $N$, $P(E_1)$ is normalized to $N/N^*$.
The vertical lines indicate positions of the curves' maxima.  For all
considered magnitudes of disorder the maximum of the local DOS is shifted
with respect to the maximum of the total DOS by about the mean spacing in
the local energy structure, $\varepsilon_{12}$.}
\label{fig2}
\end{figure}
It turns out that the structure of the exciton states in each local manifold
is very similar to the structure of the lower states of a homogeneous
(non-disordered) linear chain of length $N^*$.  In particular, the lowest
state in a manifold has a wave function without nodes within its
localization segment. Such a state can be interpreted as the {\it local}
ground state of the segment (italic is used to distinguish this state from
the true ground state, that is, the state with the lowest energy in each
realization). A {\it local} ground state carries large oscillator strength,
approximately $N^*$ times larger than that of a monomer, so that the typical
spontaneous emission rate is $\gamma^*=\gamma\,N^*$. The scaling law of the
localization length is~\cite{Malyshev01a}
\begin{equation}
N^* = 8.7\,\left(\frac{\Delta}{J}\right)^{-0.67} \ .
\label{scalingA}
\end{equation}
The energy distribution of the {\it local} ground states,
calculated as described in Ref.~\cite{Malyshev01a}, and the total
DOS are presented in Fig~\ref{fig2}. This figure shows that
almost all {\it local} ground states belong to the tail of the
DOS, as has been mentioned in the Introduction.

The second state in a manifold has a node within the localization segment
(see the states filled with black color and joined by ellipses in
Fig.~\ref{fig1}) and looks like the first {\it local} excited state of the
segment. Its oscillator strength is typically an order of magnitude smaller
than that of the {\it local} ground state. It is important to recall here
that, contrary to the eigenstates from the same manifold, the lower states
localized at different segments overlap weakly (see all states filled with
black color in Fig.~\ref{fig1}). The energies of the {\it local} ground
states are distributed within the interval $\sqrt{2}\sigma_{11}$
($\sigma_{11}$ being the average spacing between {\it local} ground states).
This interval is larger than the typical energy spacing
$\varepsilon_{12}$ between the levels in a local
manifold~\cite{Malyshev01a}. For this reason, the local energy
structure cannot be seen either in the DOS (see Fig.~\ref{fig2})
or in the linear absorption spectra (see, for instance,
Ref.~\cite{Fidder91b}). However, this structure determines the 
nonlinear optical response of the
aggregate~\cite{Minoshima94,Knoester96,Bakalis99}.

Higher states are more extended than the {\it local} states as the
localization length increases with energy (see the states filled with gray
color in Fig.~\ref{fig1}). Therefore, the higher states cannot be included
into any particular local manifold: their wave functions covers more than
one $N^*$-molecule segments. Nevertheless, as all these states are close to
the maximum of the DOS (see Fig.~\ref{fig2}), the typical energy spacing
between the higher states and the covered {\it local} states is of the order
of $\varepsilon_{12}$. Thus, the energy $\varepsilon_{12}$ is expected to be
the characteristic energy of the exciton diffusion.

It is clear from the above arguments that at temperatures $T <
\varepsilon_{12}$, it is the states from the local manifolds that determine
the exciton diffusion. Two types of hopping over
these states can be distinguished: intra-segment hopping and
inter-segment one, involving the states of the same local manifold
and those of different manifolds, respectively. As the states from
different local manifolds overlap weakly, only inter-segment hops
to adjacent segments are of importance. The disorder scaling of
the overlap integrals $I_{\mu\nu} = \sum_n \varphi_{\mu n}^2
\varphi_{\nu n}^2$ for the {\it local} states of the same and
adjacent segments was obtained in Ref.~\onlinecite{Malyshev01b}:
\begin{subequations}
\begin{equation}
I_{12} = 0.14\ \left(\frac{\Delta}{J}\right)^{0.70}
\label{scalingE}
\end{equation}
\begin{equation}
I_{\nu^\prime 1} \approx I_{\nu^\prime 2} = 0.0025\
\left(\frac{\Delta}{J}\right)^{0.75}  \ .
\label{scalingF}
\end{equation}
\label{scalingI}
\end{subequations}
\\
Hereafter, the indices $1$ and $2$ label the {\it local} states of the same
segment while those with primes label the {\it local} states of an adjacent
segment. As follows from Eq.~(\ref{scalingI}), the intra-segment overlap
integral is typically two orders of magnitude larger than the inter-segment
one. Note, that both overlap integrals scale approximately proportional to
the inverted $N^*$ (compare Eqs.~(\ref{scalingE})-(\ref{scalingF}) with
Eq.~(\ref{scalingA})). This proportionality holds for two exponential
functions extended over the length $N^*$ and separated by the distance of
the same order of magnitude, $N^*$.

The intra-segment hops do not result in the
spatial displacement of excitons. Only the inter-segment hopping gives
rise to the spatial motion. Nevertheless, we show below that
both types of hops are important for understanding the features of
the low-temperature exciton transport.

The overlap integrals between the {\it local} states of a segment
and the higher states which are extended over this segment and a
few adjacent ones (see the states filled with gray color in
Fig.~\ref{fig1}) are of the order of $I_{12}$. This fact implies
that even at $T < \varepsilon_{12}$, the indirect hops via these
higher states can be more efficient than the direct inter-segment
hops over the states of local manifolds (see below). Our calculations 
support this assumption.

\subsection{Hopping at zero temperature}
\label{Teq0}

\begin{figure}[ht!]
\includegraphics[width=\columnwidth,clip]{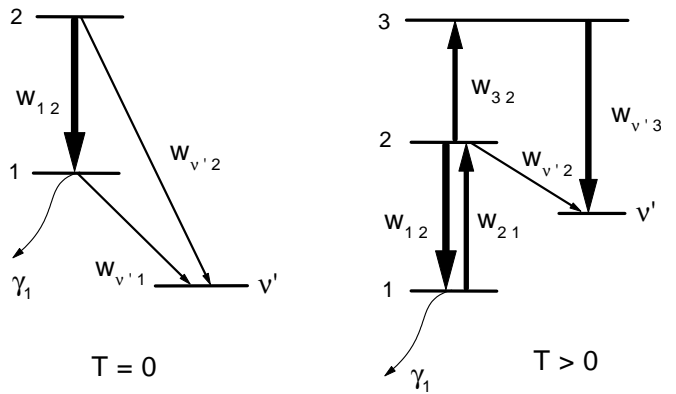}
\caption{Schematic view of exciton hopping at zero and non-zero
temperatures. The indices $1$ and $2$ label the {\it local} ground and the
first {\it local} excited states of the same segment, respectively.  The
$\nu^\prime$ state is localized at an adjacent segment. The index 3 label a
higher state, which extends over two adjacent segments. Hops are shown by
straight arrows; the arrow thickness represents magnitude of the
corespondent hopping rate. Thin wavy arrows show spontaneous emission. Its
rate is the slowest among all, which corresponds to the limit of fast
diffusion.}
\label{fig3}
\end{figure}

At zero temperature an exciton can hop only down to lower states.
Let us assume that it is in the {\it local} excited state $2$. 
Then it can either hop to the {\it local} ground state $1$ of the same
segment or to a lower state $\nu^\prime$ localized at an
adjacent segment (see Fig.~\ref{fig3}, $T=0$). Because the
intra-segment hopping is faster than the inter-segment one, first,
the exciton hops down to the {\it local} ground state $1$ with the
typical energy loss $\varepsilon_{12}$ ($\varepsilon_{12}$ being
the mean energy spacing in the local energy structure, see
Fig.~\ref{fig3}, $T=0$). From the {\it local} ground state, the
exciton can hop
only to a state $\nu^\prime$ of an adjacent segment, provided $\varepsilon_{\nu^\prime} < \varepsilon_{1}$ and the spontaneous
emission rate of the {\it local} ground state $\gamma_1$ is small
compared to the intra-segment hopping rate $W_{\nu^\prime
1}|_{T=0}$. Hereafter, such a relationship between these rates is
referred to as the limit of fast diffusion; only this limit is
considered in this work. The typical energy loss during such
sideways hop is of the order of the average spacing between {\it
local} ground states, $\sigma_{11}$ ($\sigma_{11}$ is of the order
of the J-band width). Thus, already after one such sideways hop
the exciton resides in a state in the tail of the DOS (see Fig.
\ref{fig2}). Therefore, the number of states with even lower
energies decreases dramatically, which results in a strong increase
of the typical distance to those states and decrease of the 
probability to hop further sideways. Then the exciton either
relaxes to a lower state of the same segment (if there is one) or
decays spontaneously, i.e. this type of the spatio-energetic
diffusion (towards lower energies) stops very quickly. Note that
this diffusion would manifest itself in the red shift of the
exciton emission spectrum relative to the absorption spectrum. The
experimental data shows that such red shift is either
absent~\cite{deBoer90,Fidder90} or is smaller than the J-band
width~\cite{Scheblykin00,Kamalov96}. These experimental facts
indicate unambiguously that at low temperatures, $T \ll
\sigma_{11}$, excitons make few hops before they decay due to the
spontaneous emission, as was argued in
Refs.~\onlinecite{Malyshev99b,Malyshev00,Ryzhov01}. Consequently,
the zero-temperature exciton quenching is expected to be weak
provided the concentration of quenchers is low, the case we are
interested in.

\subsection{Hopping at non-zero temperatures}
\label{Tneq0}

At non-zero but low temperatures ($0 < T \lesssim \varepsilon_{12}$), an
exciton can also hop up in energy. Consider an exciton in one of the lower
states in the tail of the DOS, e.g. in the {\it local} ground state $1$ (see
Fig.~\ref{fig2}, $T > 0$). For the reasons discussed above, first, the
exciton hops up to the first {\it local} excited state $2$ of the same
segment, provided the hopping rate for the considered temperature is larger
than the spontaneous emission rate $\gamma_1$ of the initial state $1$.
During this process the exciton typically gains the energy
$\varepsilon_{12}$. As $\varepsilon_{12}$ is of the order of
$\sigma_{11}$~\cite{Malyshev01a}, already after the first hop up the exciton
leaves the tail of the DOS (see Fig. \ref{fig2}) and, hence, it is likely to
have a lower state $\nu^\prime$ localized at an adjacent segment. A hop down
to this state with loss in energy is favorable and results in the spatial
displacement of the exciton, i.e., in the exciton diffusion. We stress that
although only sideways hops result in the spatial displacement of the
exciton, it is the initial hop up from the {\it local} ground state $1$ to
the {\it local} excited state $2$ that triggers the diffusion.

Another way for the exciton to hop sideways to the state
$\nu^\prime$ is via the higher state $3$ that overlaps well with
both states $2$ and $\nu^\prime$ (see Fig.~\ref{fig3}, $T > 0$).
As it has been mentioned, such hops compete with the sideways hops
over the {\it local} states; although the hop up to the state $3$
is thermally unfavorably, the overlap integral for this hop,
$I_{31}$, is large compared to that for an inter-segment hop,
$I_{\nu^\prime 1}$. We show later that this channel of the
diffusion becomes dominant even at relatively low temperature.

\section{Temperature dependence of the quenching rate}
\label{Numerics}

In this section, we discuss the results of numerical calculation of the
quenching rate $W_q$. We consider the initial condition where the leftmost
{\it local} ground state is excited while a single trap is located in the
center of the localization segment of the rightmost {\it local} ground
state. In this case, the exciton quenching is most affected by the
diffusion, as the created exciton has to travel over almost the whole chain
to be quenched. Thus, the exciton quenching at low concentration of traps
can be studied. The quenching rate was calculated as described in section
\ref{Model} for the parameter set corresponding to the limit of fast
diffusion and fast quenching (the latter limit is defined below).

\subsection{Numerical results}

As it was already mentioned in Sec.~\ref{Qualit}, in the limit of
fast diffusion the inter-segment down-hopping rate is large
compared to the typical spontaneous emission rate of a {\it local}
ground state:
\begin{equation}
W_{\nu^\prime 1}|_{T=0} \sim W_0\,\frac{\sigma_{11}}{J}\,
I_{\nu^\prime 1} \gg \gamma\ N^* \ . \label{fastdiffusion}
\end{equation}
If a quencher is located within the localization segment of a {\it
local} state then the typical quenching rate for this state is
$\Gamma^* = \Gamma/N^*$ [see Eq.~(\ref{Gamma})]. As we are
interested in the limit of fast quenching, this rate should be
taken larger than the typical intra-segment down-hopping
rate:
\begin{equation}
\Gamma/N^* \gg W_{12}|_{T=0} \sim W_0\,
\frac{\varepsilon_{12}}{J}\,  I_{12} \ . \label{fastquenching}
\end{equation}
This ensures that once an exciton hops to a {\it local} state of
the segment with the trap, it is quenched almost instantly.

The inequalities (\ref{fastdiffusion}) and (\ref{fastquenching})
yield the relationship between the rate equation parameters in the
limit of fast diffusion and quenching:
\begin{subequations}
\begin{equation}
W_0\ \frac{\sigma_{11}}{J}\ \frac{I_{1^\prime 2}}{N^*} \gg \gamma
\end{equation}
\begin{equation}
W_0\ \frac{\varepsilon_{12}}{J}\ I_{12}\ N^* \ll \Gamma \ .
\end{equation}
\end{subequations}
The scaling laws of the values of $\sigma_{11}$ and
$\varepsilon_{12}$  were obtained in Ref.~\cite{Malyshev01a} and
are given by
\begin{subequations}
\begin{equation}
\sigma_{11} = 0.7\,J\,\left(\frac{\Delta}{J}\right)^{1.33}\ ,
\label{scalingC}
\end{equation}
\begin{equation}
\varepsilon_{12} = 0.4\,J\,\left(\frac{\Delta}{J}\right)^{1.36}\ .
\label{scalingD}
\end{equation}
\label{scaling}
\end{subequations}
On the basis of the scaling laws (\ref{scalingI}) and
(\ref{scaling}) the parameter set was chosen so that for each
magnitude of the disorder $\Delta$ the following equalities hold:
$\Gamma^* = 10\,W_{12}|_{T=0}$ and $W_{1^\prime 1}|_{T=0} =
10\,\gamma^*$. Calculations were performed for $N=1000$ and averaged
over 100 realizations of the disorder.

\begin{figure}[ht!]
\includegraphics[width=\columnwidth,clip]{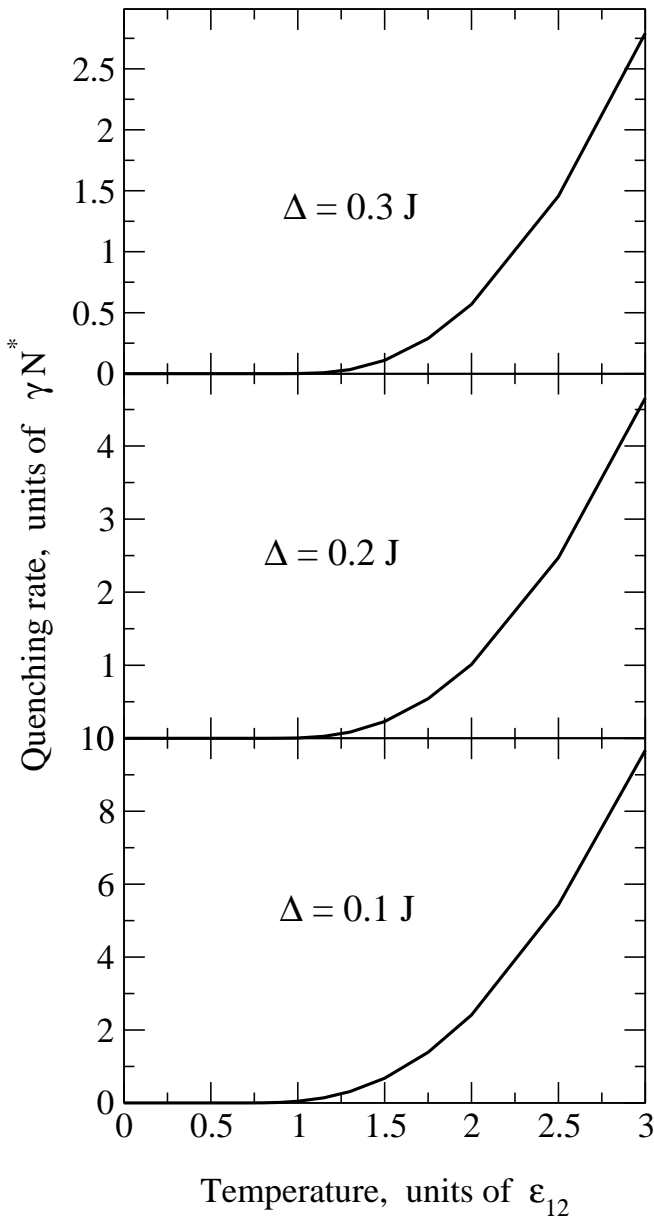}
\caption{Temperature dependence of the quenching rate $W_q$
calculated for a linear chain of the length $N = 1000$ and
different magnitudes of the disorder $\Delta$. The averaging is
performed over 100 disorder realizations. For each realization of
the disorder, the leftmost {\it local} ground state is excited,
while the only trap is located in the center of the localization
segment of the rightmost {\it local} ground state.} \label{fig4}
\end{figure}

Figure \ref{fig4} shows the temperature dependence of the
quenching  rate $W_q$ for different magnitudes of the disorder
$\Delta$. In each plot, the quenching rate is given in units of
the typical exciton radiative rate $\gamma^* = \gamma N^*$. The
temperature is given in units of the mean energy spacing in the
local energy structure $\varepsilon_{12}$. Note that both $N^*$
and $\varepsilon_{12}$ depend on $\Delta$ as described
by~(\ref{scalingA}) and~(\ref{scalingD}). Figure \ref{fig4}
demonstrates very clearly that for all considered values of
$\Delta$ at temperatures lower than $\varepsilon_{12}$ the
quenching rate is vanishing. This indicates that the diffusion at
these temperatures is not fast enough for the exciton to reach the 
quencher during its (spontaneous) lifetime: it emits a photon before it is
trapped. On the contrary, just after the temperature exceeds
approximately $\varepsilon_{12}$ the quenching becomes noticeable:
the exciton partly diffuses to the trap where it decays mostly due
to quenching. Specifically, temperature of the order of
$2\,\varepsilon_{12}$ are required for the quenching to become as
effective as the spontaneous emission: $W_q \sim \gamma^* =
\gamma N^*$.

\subsection{Discussion}
\label{Discussion}

In order to understand which states contribute most into the quenching
process it is useful to estimate the effective sideways hopping rate $W$,
which is required to reach the quenching level $W_q \sim \gamma^*$. To do
this, consider the sequence of localization segments as an effective chain
of "sites", the typical number of which is equal to the number of segments,
$N_s= N/N^*$; the mean spacing between these "sites" is $N^*$. The exciton
diffusion coefficient is then estimated as $D \sim W{N^*}^2$ (the lattice
constant is set to unity). For the quenching to be as effective as the
spontaneous decay, the exciton has to be at the position of the trap
(located on the opposite side of the chain) during the lifetime
${\gamma^*}^{-1}$, i.e., it has to diffuse over the distance $N$ during this
time. Equating the diffusion length $\sqrt{D/\gamma^*}$ to $N$, we obtain
the estimate for the required diffusion rate $W$:
\begin{equation}
W \sim \gamma^{*} (N/N^{*})^{2} \ .
\label{Weff}
\end{equation}
The localization length $N^*$ is equal to
38, 25 and 18 for $\Delta$ = 0.1, 0.2 and 0.3, respectively. Thus,
the corresponding diffusion rates $W$ are estimated as $625
\gamma^*, \, 1600\gamma^*$ and $2500 \gamma^*$ (for N=1000).
These values are about two orders of magnitude larger than the
rates of sideways hops over the {\it local} states, taken to be
$10 \gamma^*$ in all calculations. This indicates that when the
quenching rate becomes comparable to the spontaneous emission
rate, the exciton does not hop directly between the {\it local}
states of adjacent segments (with the typical rate $W_{\nu^\prime
1} \sim 10\gamma^*$). It rather hops via the higher states
that extend over more than one $N^*$-molecule segments (see the
discussion in Sec.~\ref{Qualit}). The hopping rate via such states
for $T \sim 2\,\varepsilon_{12}$ is of the order of $W_{12}$ which
is about two orders of magnitude larger than $W_{\nu^\prime 1}$ 
(see Eq.~(\ref{1Wkk'}) and the scaling laws (\ref{scalingI})).

\begin{figure}[ht!]
\includegraphics[width=\columnwidth,clip]{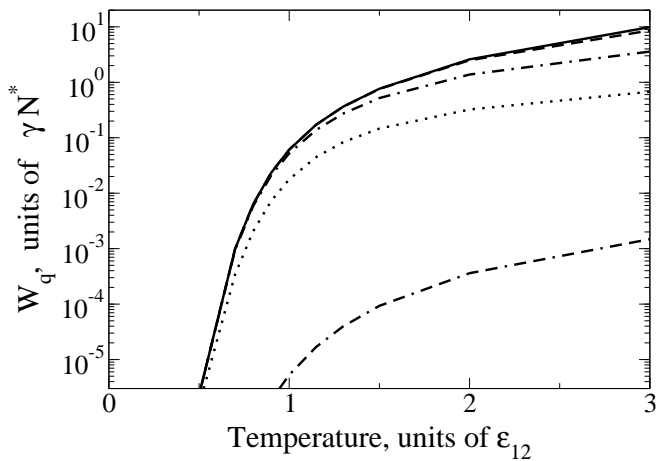}
\caption{Temperature dependence of the quenching rate $W_q$
calculated for $\Delta=0.1J$  ($N^* \approx 40$) and different numbers 
of states considered in Eq. (\ref{1tau}): 
solid line --- all $N=1000$ states, 
dashed line --- $6N/N^*=150$ states, 
dashed-dotted line --- $4N/N^*=100$ states, 
dotted line --- $3N/N^*=75$ states, 
dashed-dashed-dotted line --- $2N/N^*=50$ states.
The averaging is performed over 100 disorder realizations.
For each realization of the disorder, the
leftmost {\it local} ground state is excited, while the only trap
is located in the center of the localization segment of the
rightmost {\it local} ground state.} 
\label{fig5}
\end{figure}

In order to prove the above finding we performed calculations of the
quenching rate $W_q$ varying the number of states considered in Eq.
(\ref{1tau}). More specifically, we considered all states up to some
(variable) cut-off state. Figure~\ref{fig5} shows the results of such study
performed for $\Delta=0.1J$. As it can be seen from the figure, $W_q$
depends drastically on the number of states participating in the quenching
process. In the region where $W_q > \gamma^*$, approximately $6N/N^*$ states
are required to reach the true value of the quenching rate that is
calculated for all states (compare dashed and solid lines). Thus, at
temperatures $T \gtrsim \varepsilon_{12}$ the higher states provide the
dominant contribution into the exciton quenching process.

\begin{figure}[ht!]
\includegraphics[width=\columnwidth,clip]{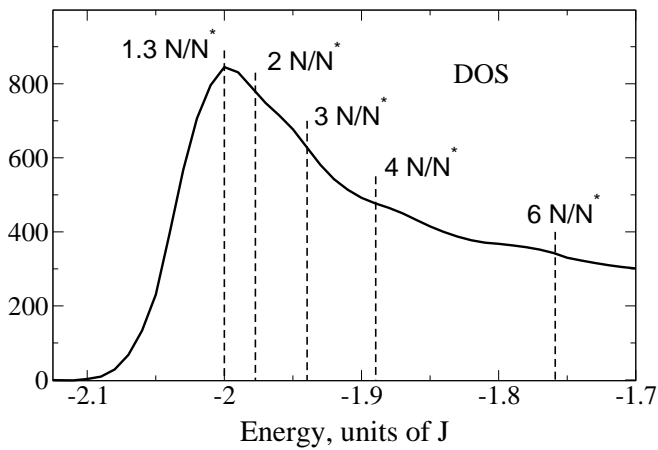}
\caption{The total DOS calculated for $N = 1000$, $\Delta=0.1J$
($N^* \approx 40$). The DOS is normalized to $N$. The vertical
lines show the maximum energies corresponding to different numbers of
states considered in the rate equation that was used to calculate the
dependencies plotted in Fig.~\ref{fig5} (in the sense that all
states lower than the specified energy are considered). Note that
the tail of the DOS is formed by $1.3 N/N^*$ states, namely, by
the states of the local manifolds ($N/N^*$ {\it local} ground
states plus $0.3N/N^*$ of the {\it local} excited states; recall
that about 30\% of the {\it local} ground states form the doublets).} 
\label{fig6}
\end{figure}

Figure~\ref{fig6} shows the regions of the DOS that correspond to different
numbers of states that were used for calculation of the data presented in
Fig.~\ref{fig5}. The higher states lie just after the {\it local} ones,
close to the maximum of the DOS (see Fig.~\ref{fig6}). Therefore, the
typical energy spacing between the {\it local} and higher states is about
$\varepsilon_{12}$. As the higher states extend over several, but
not very many, $N^*$-molecule segments (see Fig.~\ref{fig1}), the overlap
integral between these states and the covered {\it local} states is large.
These two factors ensure high hopping rate from the lower {\it local} to the
higher states. Another important point is that higher states are well
overlapped and more extended, so hops between them are typically faster and
longer than those between the {\it local} ones. Also, the higher states have
small oscillator strength, so as long as an exciton remains in these states
it does not decay radiatively. The above qualitative arguments explain the
dominant contribution of the higher states into the exciton diffusion and
quenching within the temperature range $T \gtrsim \varepsilon_{12}$.

\begin{figure}[ht!]
\includegraphics[width=\columnwidth,clip]{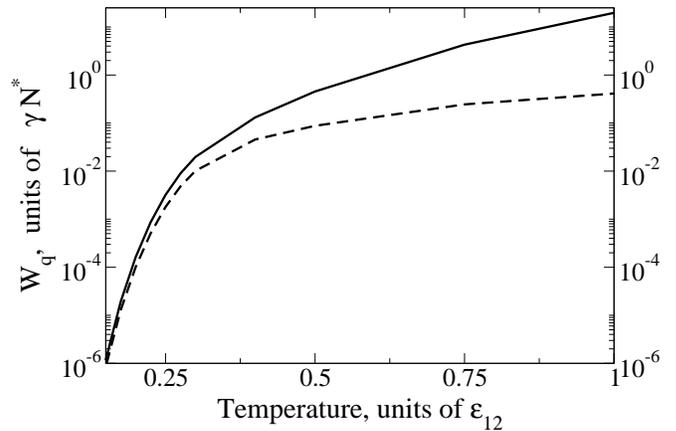}
\caption{Temperature dependence of the quenching rate $W_q$ calculated for
$\Delta=0.2J$ ($N^* \approx 25$).
Solid line --- all $N=250$ states, 
dashed line --- $2N/N^*=20$ states.}
\label{fig7}
\end{figure}

It is also seen from Fig.~\ref{fig5} that the difference between the true
value of the quenching rate $W_q$ and that calculated for a restricted
number of states decreases at lower temperatures. Figure~\ref{fig7}
demonstrates the temperature dependence of $W_q$ obtained for
$\Delta=0.2J$ ($N^*\approx 25$) and temperatures $T < \varepsilon_{12}$. The
solid line presents the dependence calculated for all $N=250$ states
considered in the rate equation, the dashed line --- that for $2N/N^*=20$
states. These $2N/N^*$ states include all the states of the local manifolds
($1.3 N/N^*$) and $0.7 N/N^*$ higher ones.

The parameters of the rate equation was set as follows: $\Gamma^* =
10\,W_{12}|_{T=0}$ and $W_{1^\prime 1}|_{T=0} = 100\,\gamma^*$. The chain
length $N=250$ is chosen so that the effective hopping rate $W \sim \gamma^*
(N/N^*)^2$ estimated as discussed above (see Eq.~(\ref{Weff})) is equal to
the rate of the direct hopping to an adjacent segment, $W_{1^\prime
1}|_{T=0}$. This yields the equation $(N/N^*)^2 = 100$ for the chain length.
This condition ensures that the diffusion over the lower states only can
provide the quenching rate $W_q \sim \gamma^*$ at $T \sim \varepsilon_{12}$.
Indeed, $W_q \approx 0.5\,\gamma^*$ at this temperature for $2N/N^*$ states. 
The most important point demonstrated by Fig.~\ref{fig7} is that below the
temperature $T_1 \approx 0.25\,\varepsilon_{12}$ the two curves deviate
slightly, which means that the contribution of the higher states into the
diffusion becomes negligible: the exciton hops mostly over the DOS tail
states. On the contrary, above the temperature $T_1$ the higher states
provide the dominant contribution to the diffusion and quenching. Note also,
that the value of the quenching rate at the critical temperature $T_1$ is
typically very small, so the experimental observation of this "regime
change" is a challenging task.

The critical temperature $T_1$ at which the higher states come into play can
be estimated by equating the typical rate of the direct sideways hopping
from a {\it local} state $2$ to an adjacent {\it local} state $1^\prime$ to
the "vertical" hopping rate from the {\it local} state $2$ to a higher state
$3$ (see Fig.~\ref{fig3}, $T>0$): $W_{1^\prime 2} = W_{32} \approx
W_{12}|_{T=0}\,\exp(\varepsilon_{12}/T_1)$. This equation yields the
temperature $T_1$:
\begin{equation}
T_1 = \frac{1}{\ln \left(I_{12}/I_{1^\prime 2}\right)}\,\varepsilon_{12} 
\approx 0.25\,\varepsilon_{12}\ .
\label{Testimate}
\end{equation}
We stress that the numerical factor $1/\ln (I_{12}/I_{1^\prime 2}) \approx
0.25$ is almost independent of the disorder, as the disorder scalings of the
overlap integrals are almost the same (see
Eqs.~(\ref{scalingE})-(\ref{scalingF})). So, the estimate
$T_1 \approx 0.25\,\varepsilon_{12}$ is universal for a wide range of
the disorder degree.

\section{Summary and concluding remarks}
\label{Concl}

In this paper, we study theoretically the peculiarities of the
low-temperature diffusion of the 1D Frenkel excitons localized by a moderate
diagonal disorder. The exciton motion over localized states is considered as
{\it incoherent} hopping. The diffusion is probed by the exciton quenching
at a point trap. We consider a single trap located at one end of the
aggregate while the exciton is created initially at the other end. In this
case the exciton has to travel over almost the whole chain to be quenched.
Under this conditions, the quenching rate carries direct information about
the diffusion length that the exciton travels over during its lifetime. The
exciton quenching is described by the rate equation with the quenching rate
being proportional to the probability of finding the exciton at the trap
site.

Both our qualitative arguments and numerical simulations show that there
exist two regimes of the exciton diffusion. At lower temperatures, those
smaller than $T_1 \approx 0.25\times$J-band-width, the exciton diffuses
mostly over {\it weakly overlapped} DOS tail states which determine the
optical response and form the J-band. This regime of diffusion is very slow;
the exciton cannot diffuse over large distance during its lifetime at these
temperatures.

At higher temperatures, the higher states come into play. The diffusion
begins to built up due to the two-step hops via higher states. This
accelerates the exciton diffusion drastically, so that an exciton can
diffuse over large distances during its lifetime. The higher states begin to
contribute dominantly to the diffusion at temperatures higher than about
$T_1$. However, the diffusion becomes really fast (in the sense that the
quenching rate becomes comparable to the spontaneous emission rate of the
aggregate) only at the temperatures of the order of the J-band width.

In Ref.~\cite{Scheblykin00}, the anomalously fast low-temperature diffusion
of Frenkel excitons in J-aggregates of THIATS was reported. The authors of
Ref.~\onlinecite{Scheblykin00} studied experimentally the exciton-exciton
annihilation and found that this effect is pronounced even at $T = 5$ K
(3.5cm$^{-1}$), while the width of J-band of THIATS J-aggregates is
82cm$^{-1}$. In order to explain the experimental data, the authors assumed
that an exciton travels over about $10^4$ dye molecules during its lifetime
to meet another exciton and annihilate. They found also that the activation
energy of the exciton diffusion was $15 K$ (10.5cm$^{-1}$) and interpreted
this energy as the typical energy spacing between the states of adjacent
localization segments.

Despite the fact that the exciton-exciton annihilation should to be treated
differently from the exciton quenching, the model we are dealing with can
easily be adapted for qualitative analysis of the exciton-exciton
annihilation: one of the two excitons can be considered as an immobile trap
for the other, while the other diffuses twice as fast. As reported in
Ref.~\cite{Scheblykin00}, the fluorescence spectrum of THIATS J-aggregates
is narrowed by approximately 26cm$^{-1}$ and red-shifted by 23cm$^{-1}$ as
compared to the absorption spectrum. These results indicate unambiguously
that excitons make sideways hops during their lifetime, i.e., the rate of
sideways hops over {\it local} states is larger than the exciton spontaneous
emission rate. Thus, the conditions for the exciton diffusion in THIATS
J-aggregates are similar to those studied in the present paper (the limit of
fast diffusion).

Discussing the above mentioned experimental data and its interpretation
presented in Ref.~\cite{Scheblykin00}, the following points can be made.
First, the typical energy spacing between the states of the adjacent
segments is of the order of the J-band width~\cite{Malyshev01b}, that is 
82cm$^{-1}$ and not 10.5cm$^{-1}$. The latter value is closer to
$0.25\times 82$cm$^{-1}$, i.e. this temperature could be related to the
temperature $T_1$, the activation energy of the faster exciton diffusion
regime. Above this temperature an exciton diffuses mostly over the higher
states and not over the DOS tail states, as it was suggested in
Ref.~\cite{Scheblykin00}.  Another point, and a more important one, is of
the quantitative nature: The typical size of the localization segment in
THIATS J-aggregates is $N^*=30$~\cite{Scheblykin00a}. In the model we are
using, this corresponds to the disorder magnitude $\Delta \approx 0.2 J$.
Our numerical data obtained for a chain of $N = 1000$ molecules demonstrates
that for this value of the disorder the exciton quenching is vanishingly
small for the temperatures $T \sim (10.5/82) \times \varepsilon_{12}$ (we
remind that $\varepsilon_{12}$ is of the order of the J-band width). In
other words, the exciton created in the leftmost {\it local} ground state
cannot diffuse over the whole chain of $1000$ monomers during its lifetime.
However, it can do so at the temperatures of the order of $T \sim
\varepsilon_{12} \sim 82$cm$^{-1}$. Thus, understanding the fast
low-temperature diffusion in the aggregates of THIATS dye molecules, observed
in Ref.~\cite{Scheblykin00}, still remains an open question.

\acknowledgments

This work was supported by the DGI-MCyT (Project MAT2000-0734).
A.~V.~M. and F.~D.~A. acknowledge support from CAM (Project
07N/0075/2001). V.~A.~M. acknowledges support from MECyD (Project
SAB2000-0103) and through a NATO Fellowship.

\end{document}